%% file: _main.tex
\documentclass[10pt,conference]{IEEEtran}
\usepackage{amsmath,amssymb,amsfonts}
\usepackage{algorithmic}
\usepackage{balance}
\usepackage{graphicx}
\usepackage{textcomp}
\usepackage{xcolor}
\usepackage[T1]{fontenc} % for < symbol
\usepackage{threeparttable}

\def\BibTeX{{\rm B\kern-.05em{\sc i\kern-.025em b}\kern-.08em
    T\kern-.1667em\lower.7ex\hbox{E}\kern-.125emX}}

\begin{document}

\title{Improving Software Engineering Team Communication Through Stronger Social Networks}

\author{\IEEEauthorblockN{
April Clarke}
\IEEEauthorblockA{\textit{Computer Science} \\
\textit{and Software Engineering} \\
\textit{University of Canterbury}\\
Christchurch, New Zealand \\
april.clarke@pg.canterbury.ac.nz}
\and
\IEEEauthorblockN{
Tanja Mitrović}
\IEEEauthorblockA{\textit{Computer Science} \\
\textit{and Software Engineering} \\
\textit{University of Canterbury}\\
Christchurch, New Zealand \\
tanja.mitrovic@canterbury.ac.nz}
\and
\IEEEauthorblockN{
Fabian Gilson}
\IEEEauthorblockA{\textit{Computer Science} \\
\textit{and Software Engineering} \\
\textit{University of Canterbury}\\
Christchurch, New Zealand \\
fabian.gilson@canterbury.ac.nz}
}

\maketitle

\input{abstract}
\input{introduction}
\input{related_work}
\input{method}
\input{results}
\input{discussion}
\input{conclusion}

\clearpage
\balance
\bibliographystyle{IEEEtran}
\bibliography{IEEEabrv,refs.bib}

\end{document}

%% file: abstract.tex
\begin{abstract}
Students working in teams in software engineering group project often communicate ineffectively, which reduces the quality of deliverables, and is therefore detrimental for project success. An important step towards addressing areas of improvement is identifying which changes to communication will improve team performance the most. We applied two different communication analysis techniques, triad census and socio-technical congruence, to data gathered from a two-semester software engineering group project. Triad census uses the presence of edges between groups of three nodes as a measure of network structure, while socio-technical congruence compares the fit of a team's communication to their technical dependencies. Our findings suggest that each team's triad census for a given sprint is promising as a predictor of the percentage of story points they pass, which is closely linked to project success. Meanwhile, socio-technical congruence is inadequate as the sole metric for predicting project success in this context. We discuss these findings, and their potential applications improve communication in a software engineering group project.
\end{abstract}

%% file: introduction.tex
\section{Introduction}
Communication and balanced contributions between team members are widely recognised as key components of effective teamwork, and are crucial for the success of software engineering teams~\cite{curtisFieldStudySoftware1988,hoeglTeamworkQualitySuccess2001, mensahImpactTeamworkQuality2024}. However, teams are often ineffective at these skills~\cite{krautCoordinationSoftwareDevelopment1995}, leading to reduced productivity and effectiveness~\cite{r.noelExploringCollaborativeWriting2018,strodeTeamworkEffectivenessModel2022a}, which is detrimental for project success. In particular, teams identify a lack of knowledge sharing within the team and communication breakdowns as sources of issues, especially in times of stress~\cite{salasHowCanYou1997}.

There are often differences between members' participation in team discussions. More vocal team members tend to dominate discussions, causing the team to miss knowledge from quieter team members that could be used to get the best solutions~\cite{hoeglTeamworkQualitySuccess2001}. Knowledgeable team members who are also skilled communicators tend to become ``communication focal points'', and gather knowledge through the process of educating their teams about the application domain~\cite{omalleyAnalysisSocialNetworks2008}. These team members are recognised by their teammates as being exceptional contributors to their teams' success~\cite{curtisFieldStudySoftware1988}. However, this behaviour leads to reduced performance at the team level~\cite{sparroweSocialNetworksPerformance2001}. In well performing teams, members tend to know other members' expertise~\cite{r.noelExploringCollaborativeWriting2018}, and so are able to coordinate knowledge without depending on these ``communication focal points'' to broker information. These teams with better coordination skills show not only better performance, but also higher team member satisfaction than their counterparts~\cite{mensahImpactTeamworkQuality2024,seersTeammemberExchangeQuality1989}, indicating that improving communication and coordination in teams provides benefits at both the individual and team levels. 

To identify communication behaviours that can negatively affect team productivity, we can use social network analysis~\cite{otteSocialNetworkAnalysis2002}. Social networks represent relationships between entities, where each node in the network is an entity, and each edge between two nodes represents a relationship between them. Through analysing the characteristics of these networks and identifying communication patterns that occur in less and more productive teams, we can attempt to address negative patterns and foster positive patterns. 

While there is extensive research on the characteristics of effective teams and the importance of communication~\cite{mathieuTeamEffectiveness199720072008}, to our knowledge there is little work regarding \textit{how} we can improve social networks in team discussions based on social network analysis. We propose that social network analysis can be used to identify areas of improvement in teams' communication, and these improvement areas can inform interventions that help teams strengthen their communication, leading to better performance. In the context of software engineering education, this is valuable for helping students identify weaknesses and track their progress. It may also provide educators with early indicators of teams that need additional support to develop strong communication.  

In this paper, we first present prior work that informs our analysis of team communication in Section~\ref{rel_work}, followed by the method we used to apply these analysis techniques in Section~\ref{method}. In Section~\ref{results}, we discuss our results, and in Section~\ref{sec:discussion}, we explain the implications and limitations of our findings. Finally, Section~\ref{conclusion} concludes and discusses applications of the results in future work.

%% file: related_work.tex
\section{Background and Related Work}
\label{rel_work}
Social network analysis is a means to understand the relationships between the nodes in a network. In the context of software engineering, this can be an analysis of developer social networks. Developer social networks model relationships between developers based on data like bug reports, files, or projects developers have in common~\cite{zhangUnderstandingStudentTeachers2022}. 

% Some existing work relates the characteristics of social networks, like the presence of certain triad types, to team and individual performance.

% Betweenness centrality represents how frequently a node is found in the shortest path between another two nodes. O'Malley and Marsden~\cite{omalleyAnalysisSocialNetworks2008} found that nodes with higher betweenness centrality acted as communication brokers in networks, and tended to control relationships between other nodes. Once this anti-pattern has been identified in a team, a corrective intervention could encourage task allocations that spread knowledge across more teammates.

Triads, which are groups of three nodes, can also be useful for describing the structure of teams' social networks~\cite{zverevaTriadCensusUsage2016}.  The triad census of a network is a set containing the frequencies of each triad type. Triads are valuable as they represent different relationship structures. The presence of many triads with zero edges indicates that many team members are not communicating with each other, while many triads with two edges indicates that team members often communicate in pairs, and rarely communicate with members outside these pairs. This is considered particularly meaningful when analysing networks with few nodes~\cite{frankSurveyStatisticalMethods1981}, making it an appropriate characteristic for analysing teams of software developers. Finding the triad census of a network also contributes to calculating other network characteristics that can be used to better understand social relationships, like transitivity~\cite{newmanClusteringPreferentialAttachment2001}, so this is a useful first step. 

While these network characteristics give some insight into the structure of networks, more sophisticated analysis techniques have been developed to give insight into specific team behavioursmple, Socio-technical congruence (STC) , which are identified through file changes in the project's repository~\cite{cataldoSociotechnicalCongruenceFramework2008}. If two developers are assigned to tasks that require them to work on the same file, or files that are recognised as dependent on each other, then STC indicates that there is a need for the developers to coordinate with each other. The result of applying STC to a project is a matrix representing the degree to which each developer needs to coordinate with each of their teammates. 

In professional software teams, STC scores were positively correlated with productivity on both a team and individual level. Further, more productive team members  improved their STC scores over time~\cite{cataldoIdentificationCoordinationRequirements2006}. Conversely, MacKellar's examination of students' STC scores in a team project showed that students' STC scores did not increase over time~\cite{b.k.mackellarAnalyzingCoordinationStudents2013}. While the sample size for this study was small, the results suggest that STC score trends may be less reflective of students' productivity than industry developers. Sierra et al. suggest this is a result of students being inexperienced with version control systems~\cite{sierraSystematicMappingStudy2018}. However, the student with the highest student STC score worked on a central piece of the system and frequently mentored other students, indicating that we may identify team members who act as communication focal points through this type of analysis.

While STC is effective for identifying areas for improvement in teams' communication in professional software development teams, student teams sometimes behave differently. However, Sierra et al. identified a lack of existing work in this area~\cite{sierraSystematicMappingStudy2018} in 2018, and to our knowledge there has not been extensive work on this since then. So, it is necessary to evaluate the suitability of STC in the context of software engineering project courses. There is also little prior work on triad census as a measure of communication in software teams. In this paper, we contribute to our understanding of the suitability of STC and triad census for identifying communication areas of improvement in student teams, by evaluating them in the context of a software engineering project course.

% Differences in the work structure, team members' prior experiences, and tools they are expected to use all affect how well a metric, or combination of metrics, can accurately reflect teams' areas of improvement. For this reason, it is necessary to investigate the suitability of different analysis techniques in identifying areas for improvement in the context of Agile software engineering project courses.

% MacKeller proposes the implementation of a tool that can recommend teammates students should communicate with for a given task, and notify students when their work has been affected by a teammate's. 

%% file: method.tex
\section{Method}
\label{method}

\subsection{Context of Study}
We conducted this research in the context of SENG302, a 300-level two-semester software engineering group project at the University of Canterbury. Students work in teams of 6-8 members to develop a software application while following the Scrum framework. SENG302 is divided into seven Sprints, which are each 2-3 term weeks long, although some sprints span term breaks. Students attend regular formal sessions, including the Sprint Planning, Daily Scrums (which occur bi-weekly in this case), Sprint Review, demonstration, and Sprint Retrospective sessions. In 2023, students submitted peer-feedback for each of their teammates and a self-reflection each Sprint. These discussed what each person did well, what they could do better, and actions they could take to improve. Peer-feedback also included ratings on a Likert scale from one to five for each teammates' contribution towards the project through skills like testing and communication. Communicating technical knowledge and opinions, and with teammates in a professional environment is included in the course learning outcomes, so students are expected to improve this skill during the year.

In 2023, there were 77 students enrolled in the course, who were split into 10 teams. 
% 19 students identified as women, and 58 identified as men. Students were between 19 and 31 years old, with most of the cohort 20 or 21. 
All students had prior knowledge of programming concepts, including introductory software engineering and databases courses.

% 47 students were enrolled in software engineering, while 30 were enrolled in computer science degrees. 

To apply the analysis techniques we identified from prior work to this cohort, we collected data from four sources. From Slack, we collected messages sent in public channels in each team's workspace. We retrieved code contribution and merge request (MR) data from a locally hosted instance of GitLab. We collected work logs and peer-feedback from the project management tool, Scrumboard~\cite{minish}, which is developed locally. 

Finally, we took teams' Sprint results in the form of delivered story points per sprint, and the overall team performance score following the course marking schedule, e.g., knowledge of Scrum, quality of code, documentation.

%We wrote Python scripts for data processing with Pandas, NetworkX, SciPy, and matplotlib. We also developed network visualisations in the project management tool.

We received ethics approval from the University of Canterbury Human Research Ethics Committee to undertake this research.

\subsection{Socio-Technical Congruence}
\begin{figure}
    \centering
    \includegraphics[width=1\linewidth]{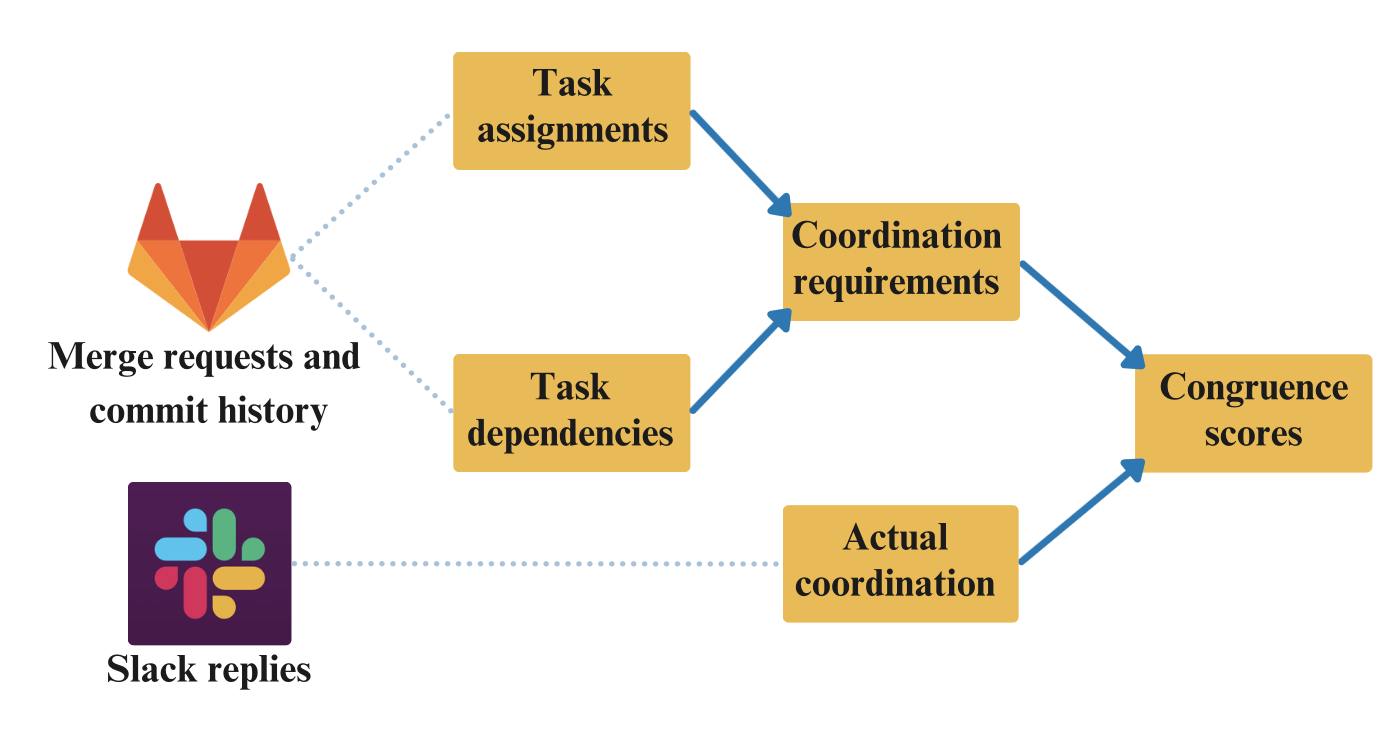}
    \caption{The process for calculating socio-technical congruence}
    \label{fig:socio-tech-flowchart}
\end{figure}

We followed the method for calculating STC scores as proposed by Cataldo et al.~\cite{cataldoIdentificationCoordinationRequirements2006}, illustrated in Fig.~\ref{fig:socio-tech-flowchart}. 

\begin{enumerate}
\item The task assignments matrix maps students to the MRs they committed to. We retrieved students' commit histories, and the commits in each MR, and combined these to construct this matrix.

\item The task dependencies matrix maps each MR to other MRs that have any files in common. %We omitted MRs from the test branch to the production branch, as they would add noise to the dependencies. While including these MRs would ensure we consider commits made directly to the test branch, these are rare, and these MRs typically contain entire stories or multiple stories of functionality. Including this many commits in the same MR would create task dependencies between all files, leading to every team member having a coordination requirement with every other team member, which would be less accurate.

\item We combined the task assignments and task dependencies matrices to produce a coordination requirements matrix, which reflects which teammates a student is expected to need to communicate with in a given week. For example, if student X contributed to MR X, and student Y contributed to MR Y, and MR X and MR Y have files in common, then student X and student Y have a need to coordinate.

\item We used communication utterances from public channels in each team's Slack workspace to construct the actual coordination matrix. We considered communication to have occurred from student X to student Y when student X sent a message in a thread started by student Y. 

\item We calculated each students' STC score for each week based on the communication requirements matrix and actual communication matrix. For each student, we created a list where each value represents whether they communicated with another teammate as expected or not. Each value is one if there was a communication requirement that was fulfilled, zero if there was a communication requirement that was not fulfilled, and null if there was no communication requirement. A student's STC score for the week is the mean of these values. Finally, we obtained each team's mean STC score for the year by taking the mean of the team members' STC scores for that week, excluding null values.
\end{enumerate}

\subsection{Triad Census}
\begin{figure}
    \centering
    \includegraphics[width=0.4\linewidth]{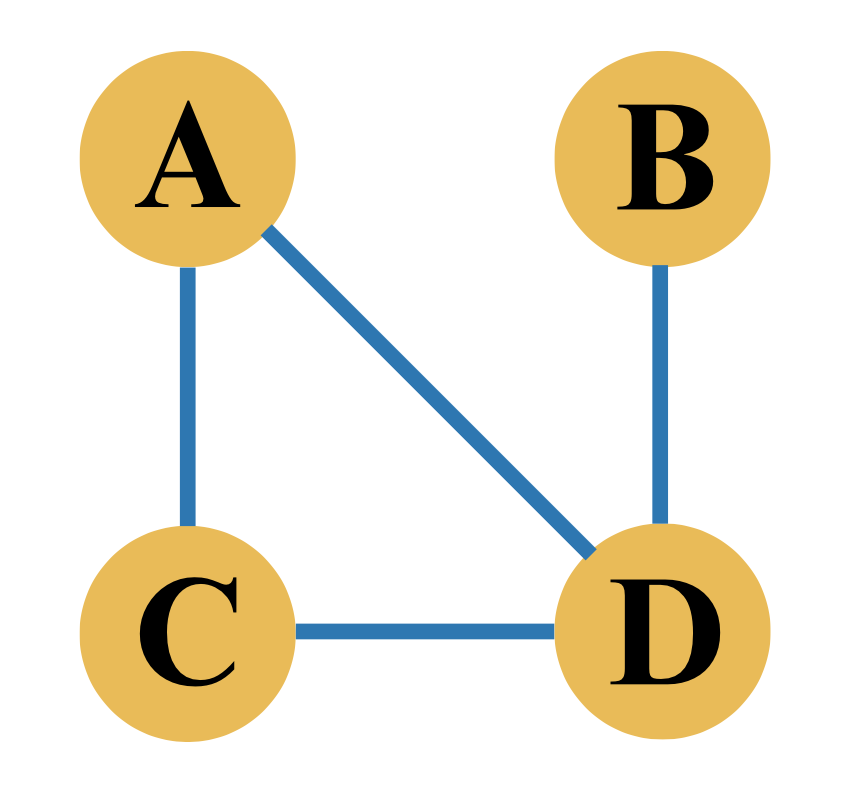}
    \caption{An example network}
    \label{fig:census_example}
\end{figure}

We constructed undirected networks for each team, with each network representing their written communication on Slack during a given Sprint. Each node in a network is a student, and each edge represents communication from one student to another. If student X sent a message in a thread created by student Y for the given Sprint, then there is an edge between student X and student Y. We then found the triad census for each week and Sprint, which involved finding the number of each triad type in the network for the given time period. In an undirected network, there are four possible triad types, which are determined by the number of edges between nodes in the triad: zero, one, two, or three edges. For example, the network in Fig.~\ref{fig:census_example} has four nodes, and four distinct triads: ABC, ABD, ACD, and BCD. These triads have one, two, three, and two edges, respectively, so there are zero triads with zero edges, one triad with one edge, two triads with two edges, and one triad with three edges. Therefore, the triad census is (0, 1, 2, 1).

As teams were different sizes, the nominal count of triads differed between teams' networks, so we converted the triad census to the relative triad census. We did this by finding the relative frequency of each triad type for each Sprint, and the mean relative frequency of each type of triad for each week of each Sprint. For our example in Fig.~\ref{fig:census_example}, the relative frequency of triads in the census (0, 1, 2, 1) is (0, 0.25, 0.5, 0.25).

We then calculated the Pearson correlation for each team's triad census for each Sprint with the percentage of committed story points the team passed in that Sprint, as well as the teams' performance score.

%% file: results.tex
\section{Results}
\label{results}
\subsection{Socio-technical congruence}
\begin{figure}[ht]
    \centering
    \includegraphics[width=1\linewidth]{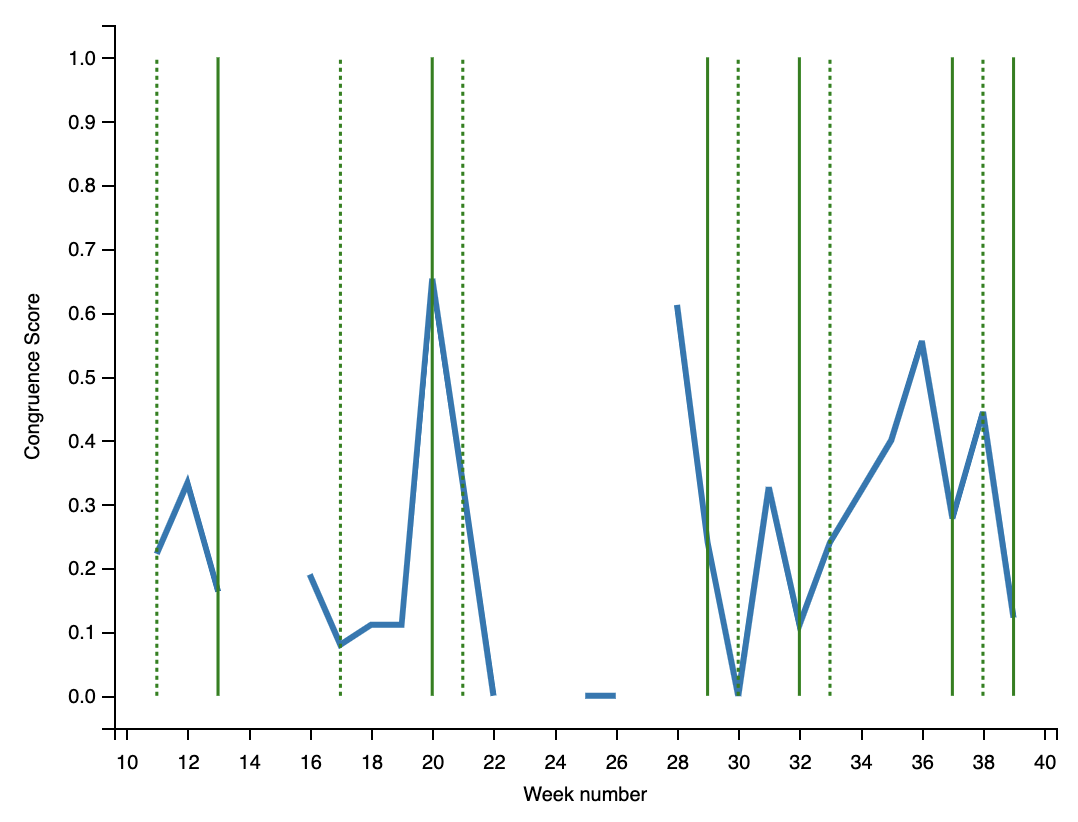}
    \caption{Mean STC scores for a team for each week of the project (Sprint 2 - Sprint 5)}
    \label{fig:congruence_over_year}
\end{figure}

We plotted each of the teams' mean STC scores for each week of the year, as shown in Fig.~\ref{fig:congruence_over_year}. Sprint 1 is excluded, as teams are split in half and work on separate projects for this initial Sprint for on-boarding purposes. In Fig.~\ref{fig:congruence_over_year}, dotted and solid vertical lines indicate Sprint start and end weeks, respectively. Note that the gap mid-year is due to exam and break period, when students do not work on the project.

Plotting the line of best fit on each team's graph showed that five out of ten teams' STC scores tended to increase over the year, while the other five decreased over the year. A Mann-Whitney U Test comparing the set of teams whose STC scores increased over the year with those who decreased over the year showed no significant difference in the number of stories passed ($U = 15.00$, $p = .68$). This aligns with MacKeller's findings that the scores of student teams in a semester-long group project did not converge~\cite{b.k.mackellarAnalyzingCoordinationStudents2013}. 

% However, MacKellar found that analysing individual students' congruence scores may be valuable for identifying students who act as knowledge brokers.

We increased the granularity of analysis by then taking the mean STC score for each team for each Sprint. The Pearson correlation coefficients for STC scores with the percentage of story points teams passed each Sprint mean rating team members gave their peers for communication in the peer-feedback for that Sprint, are shown in Table~\ref{tbl:pearson}. 

\input{tables/pearson}

The correlation between the percentage of story points passed in a Sprint and the mean peer-feedback scores for communication is significant, with a p-value of 0.003. This indicates that teams pass a lower proportion of their committed stories in Sprints where team members perceive their teammates' communication as being less effective.

We cannot intervene on a Sprint that has finished, so the correlation between the mean rating students give each other and the percentage of points they deliver is not as useful as a metric that can inform interventions during a given Sprint. However, it reinforces the need for an intervention to mitigate communication issues before the product quality is impaired. Still, a correlation between the mean rating of communication in Sprint \textit{n}, and the proportion of stories passed in Sprint \textit{n+1} would be useful. While the Pearson correlation coefficient for this is insignificant ($r = 0.26$, $p = .063$), the correlation between the mean communication score and the final team score is significant and moderate ($r = 0.33$, $p < .019$).

% \begin{itemize}
%     \item extremes not necessarily more productive
% \end{itemize}

% One of the teams did not use merge requests for the first half of the year, so we excluded this team from the analysis of congruence scores over the year.

% To investigate whether teams use in-person communication as an alternative to written online communication, we plotted the time they spent pair programming each week over their mean weekly congruence scores, as shown in figure~\ref{fig:congruence_and_pairing}. From this, we observe that weeks with low congruence scores are sometimes filled by higher pair programming, indicating that teams may use this as an alternative coordination technique. This may be pair programming between those who need to coordinate, or an indicator that the team is co-locating more, and thus uses informal communication during their co-locations to resolve coordination issues. 

% It is also notable that teams engage in less pair programming during the first week of each Sprint, likely due to the week being partially devoted to planning, and the Sprint deadline being further away than in other weeks. This indicates that students may use pair programming more in times of stress, or when they have implemented more of each feature.
% \begin{figure}
%     \centering
%     \includegraphics[width=1\linewidth]{images/Screenshot 2024-09-13 at 9.51.09 AM.png}
%     \caption{Time spent pair programming each week and mean congruence scores for a SENG302 team}
%     \label{fig:congruence_and_pairing}
% \end{figure}

\subsection{Triad census}
After calculating the triad census for each team and each Sprint, we correlated the relative frequency of the number of edges in each triad with the percentage of story points passed and performance score the team received in each Sprint.
Tables~\ref{tbl:triad_census} and \ref{tbl:triad_census_avg} show the results of this, with significant results denoted by asterisks. We note that the only significant correlation is a moderate negative correlation between the score a team receives for a given Sprint, and the relative frequency of triads with one edge. Teams with many of these triads have many pairs of members where neither member is communicating with other members of the team. This may be indicative of pair silos, where members form knowledge silos by working with only one teammate.

\input{tables/triad_census}

We then looked for anomalies in teams' communication methods, where we identified two teams that did not follow the norms based on the data presented in Table~\ref{tbl:team_stats}. We note that team G had the highest mean STC score for the year, and yet passed the least stories of any team. This may be an indicator of low quality communication. The second anomalous team, Team J, engaged in little written communication relative to the dependencies between team members' work, as shown by their low mean STC score for the year. However, their work logs reported that they engaged in more hours of pair programming than any other team, indicating more in-person communication.

\input{tables/team_stats}

Tables~\ref{tbl:triad_census_sans_anomalies} and \ref{tbl:triad_census_avg_sans_anomalies} show the correlation results with the two anomalous teams excluded. The first significant result is the correlation between the scores teams received for a given Sprint and their relative frequency of triads with one edge, which remains significant, and the coefficient has strengthened from ($r = -0.30$, $p = .033$) to ($r = -0.42$, $p = .006$). Further, the correlation between the percentage of committed story points teams pass and their relative frequency of triads with one edge is now significant with a negative correlation coefficient of ($r = -0.38$, $p = .008$). Finally, the percentage of committed story points passed is positively correlated with the relative frequency of closed triads, where each node is connected to both other nodes in the triad, ($r = 0.29$, $p = .046$).

\input{tables/triad_census_sans_anomalies}

%% file: tables/pearson.tex
\begin{table}
    \caption{Pearson Correlation Coefficients}
    \begin{center}
         \label{tbl:pearson}
        \begin{tabular}{|p{0.31\columnwidth} | p{0.3\columnwidth} | p{0.22\columnwidth}|} 
            \hline &\textbf{Mean Sprint STC Score}&\textbf{\% Story Points Passed}
            \\ \hline
            \textbf{\% Story Points Passed} & 0.109 & -
            \\ \hline
            \textbf{Mean Peer Feedback Communication Score} & 0.156 & 0.377$^{*}$
            \\ \hline
            \multicolumn{3}{l}{$^{*}$p < .05}
            \\ \multicolumn{3}{l}{$^{**}$p < .01}
        \end{tabular}
    \end{center}
\end{table}

%% file: tables/triad_census.tex
\begin{table}
\caption{Pearson Correlations Between Relative Frequencies of Triad Types and Sprint Performance}
    \begin{center}
         \label{tbl:triad_census}
        \begin{tabular}{|p{0.15\columnwidth} | p{0.37\columnwidth} | p{0.2\columnwidth}|} 
            \hline
            & \textbf{\% Passed Story Points} & \textbf{Team Score} 
            \\ \hline
            0 edges & 0.02 & -0.01
            \\ \hline
            1 edge & -0.22 & -0.30$^{*}$
            \\ \hline
            2 edges & -0.01 & -0.04
            \\ \hline
            3 edges & 0.12 & 0.19
            \\ \hline
        \end{tabular}  
    \end{center}
\end{table}

\begin{table}
    \caption{Pearson Correlations Between Mean Weekly Relative Frequencies of Triad Types and Sprint Performance}
    \begin{center}
        \label{tbl:triad_census_avg}
        \begin{tabular}{|p{0.15\columnwidth} | p{0.37\columnwidth} | p{0.2\columnwidth}|} 
            \hline
            & \textbf{\% Passed Story Points} & \textbf{Team Score} 
            \\ \hline
            0 edges & -0.20 & -0.23
            \\ \hline
            1 edge & 0.13 & 0.25
            \\ \hline
            2 edges & 0.10 & 0.08
            \\ \hline
            3 edges & 0.28 & 0.20
            \\ \hline
        \end{tabular}
    \end{center}
\end{table}

% Other header options:
% - Correlations Between Average Weekly Relative Frequencies of Triad Types and Sprint Performance
% - Correlations Between Average Weekly Frequencies of Triad Types and Sprint Performance
% - Correlations Between Average Weekly Triad Frequencies and Sprint Performance

%% file: tables/team_stats.tex
\begin{table}
    \caption{Team summaries}
    \begin{center}
        \label{tbl:team_stats}
        \begin{tabular}{| p{0.06\columnwidth} | p{0.26\columnwidth} | p{0.23\columnwidth} | p{0.2\columnwidth} |} 
            \hline 
            \textbf{Team}&\textbf{Pair Programming Hours}&\textbf{Mean STC Score}&\textbf{Stories Passed}
            \\ \hline
            A & 169 & 0.18 & 46
            \\ \hline
            B & 180 & 0.28 & 35
            \\ \hline
            C & 261 & 0.19 & 28
            \\ \hline
            D & 292 & 0.23 & 46
            \\ \hline
            E & 343 & 0.18 & 40
            \\ \hline
            F & 426 & 0.24 & 53
            \\ \hline
            G & 434 & 0.53 & 23
            \\ \hline
            H & 436 & 0.45 & 34
            \\ \hline
            I & 493 & 0.22 & 36
            \\ \hline
            J & 602 & 0.11 & 29
            \\ \hline
        \end{tabular}
    \end{center}
\end{table}

%% file: tables/triad_census_sans_anomalies.tex
\begin{table}
\caption{Correlations Between Relative Frequencies of Triad Types and Sprint Performance Excluding Anomalies}
    \begin{center}
        \label{tbl:triad_census_sans_anomalies}
        \begin{tabular}{|p{0.15\columnwidth} | p{0.37\columnwidth} | p{0.2\columnwidth}|} 
            \hline
            & \textbf{\% Passed Story Points} & \textbf{Team Score} 
            \\ \hline
            0 edges & -0.06 & 0.02
            \\ \hline
            1 edge & -0.38$^{*}$ & -0.42$^{**}$
            \\ \hline
            2 edges & 0.10 & 0.00
            \\ \hline
            3 edges & 0.20 & 0.25
            \\ \hline
        \end{tabular}
    \end{center}
\end{table}

\begin{table}
    \caption{Correlations Between Mean Weekly Relative Frequencies of Triad Types and Sprint Performance Excluding Anomalies}
    \begin{center}
        \label{tbl:triad_census_avg_sans_anomalies}
        \begin{tabular}{|p{0.15\columnwidth} | p{0.37\columnwidth} | p{0.2\columnwidth}|} 
            \hline
            & \textbf{\% Passed Story Points} & \textbf{Team Score} 
            \\ \hline
            0 edges & 0.23 & -0.25
            \\ \hline
            1 edge & 0.11 & 0.23
            \\ \hline
            2 edges & 0.11 & 0.09
            \\ \hline
            3 edges & 0.29$^{*}$ & 0.26
            \\ \hline
        \end{tabular}
    \end{center}
\end{table}

%% file: discussion.tex
\section{Discussion and Limitations}
\label{sec:discussion}
\subsection{Implications}
We have established that the percentage of story points teams pass in a given Sprint is significantly affected by their communication, measured as the communication scores team members give each other in Sprint peer-feedback, so we expect that an accurate measure of communication in a Sprint would also be correlated with the percentage of story points passed. 

As there was no significant correlation between the trend in teams' STC score and their percentage story points passed or peer-feedback communication scores, \textbf{we do not consider STC to be adequate as the sole indicator of communication areas of improvement for student teams}. However, it may still provide some insights when paired with other analysis techniques, although we have not explored this yet.  

Triad census is a promising indicator of teams' communication areas of improvement, with a significant, moderate, and positive correlation with the score teams receive for a given Sprint. Identifying and excluding anomalous teams further strengthens this correlation, and creates a significant correlation with the percentage of story points passed. The presence of these anomalous teams indicates that teams' communication is not consistently accurately represented by written communication, although this was adequate for most teams. More work is needed to strengthen our representation of communication to either include the teams that are not captured by our current implementation, or autonomously identify them. However, \textbf{triad census remains suitable for identifying teams' communication areas of improvement in most cases we observed.}

\subsection{Limitations}
There are limitations of our implementations of these analysis techniques. For STC, we expect communication for a merge request to occur in the week it was created in. However, this is inaccurate for commits made in weeks before the MR's creation, as communication for this work may occur when the commit was made, instead of when the MR was made. This could be mitigated by basing the task dependency matrix on files committed during a given time frame around each commit's creation.

For both STC and triad census, we retrieve communication from public channels in Slack, ignoring the content of messages, and assuming Slack is an adequate approximation of all communication mediums. Future work could combine this written communication with in-person communication like pair programming. We rely on the team using threads properly, as we did not include replies to messages outside threads. %Some teams in the 2024 cohort rarely or never use threads, so applying interventions based on communication threads would be an ineffective solution for improving communication in these teams. 
A potential mitigation of this is conversation disentanglement~\cite{elsnerYouTalkingMe2008}, which recreates threads where replies to a message are not in a thread. Finally, we only access communication in public channels, so communication in private channels and elsewhere is not represented.

% , and investigate whether congruence scores vary by the day of the week, which may reflect in-person communication occurring during co-location sessions and meetings.

%% file: conclusion.tex
\section{Conclusions and Future Work}
\label{conclusion}
% Notably, there is a moderate negative correlation between the frequency of triads with one edge and the score the team received for the Sprint, indicating that teams with many people communicating in pairs and not engaging in threads with others in the team tend to receive worse scores in the team assessment.
We evaluated the suitability of STC and triad census for identifying communication areas of improvement in teams of software engineering students. Our results indicate that triad census can be used to predict team performance, while STC does not indicate performance in this context.

We intend to use these findings to inform the design of features that we will add to Scrumboard to help students identify and improve where they can improve their contributions to the team. We will then evaluate the effects of these features in the context of SENG302. To do this, we will first implement a set of features that assist students in reflecting on their contributions, like a visualisation of their teams' social network, or their work consistency. Then, we will run a brief pilot study with past students, and improve the features based on their feedback. Finally, we will deploy the features for semester two of the course, and analyse the effects on students and their delivered products compared to previous cohorts, considering students' characteristics and interactions with the features. 

\section{Data Availability}
We have provided anonymised data for the Slack exports, average peer-feedback ratings for each sprint, scores and points for each team by sprint, and total time each team spent pair programming. Note that the peer-feedback ratings omit sprint 1, as we did not use these due to the teams being split into sub-teams.